# Acoustic Phonon Lifetimes Limit Thermal Transport in Methylammonium Lead Iodide

Aryeh Gold-Parker[a,b], Peter M. Gehring[c], Jonathan M. Skelton[d], Ian C. Smith[a], Dan Parshall[c], Jarvist M. Frost[e], Hemamala I. Karunadasa[a], Aron Walsh[f,g], Michael F. Toney[b,1]

[a] *Department of Chemistry, Stanford University, Stanford, CA, USA*
[b] *SLAC National Accelerator Laboratory, Stanford Synchrotron Radiation Lightsource, Menlo Park, CA, USA*
[c] *National Institute of Standards and Technology, NIST Center for Neutron Research, Gaithersburg, MD, USA*
[d] *Department of Chemistry, University of Bath, Bath, UK*
[e] *Department of Physics, Kings College London, London, UK*
[f] *Department of Materials, Imperial College London, London, UK*
[g] *Department of Materials Science and Engineering, Yonsei University, Seoul, Korea*

[1]To whom correspondence should be addressed. Email: mftoney@slac.stanford.edu

## Abstract

Hybrid organic-inorganic perovskites (HOIPs) have become an important class of semiconductors for solar cells and other optoelectronic applications. Electron-phonon coupling plays a critical role in all optoelectronic devices, and although the lattice dynamics and phonon frequencies of HOIPs have been well studied, little attention has been given to phonon lifetimes. We report high-precision measurements of acoustic phonon lifetimes in the hybrid perovskite methylammonium lead iodide (MAPI), using inelastic neutron spectroscopy to provide high energy resolution and fully deuterated single crystals to reduce incoherent scattering from hydrogen. Our measurements reveal extremely short lifetimes on the order of picoseconds, corresponding to nanometer mean free paths and demonstrating that acoustic phonons are unable to dissipate heat efficiently. Lattice-dynamics calculations using ab-initio third-order perturbation theory indicate that the short lifetimes stem from strong three-phonon interactions and a high density of low-energy optical phonon modes related to the degrees of freedom of the organic cation. Such short lifetimes have significant implications for electron-phonon coupling in MAPI and other HOIPs, with direct impacts on optoelectronic devices both in the cooling of hot carriers and in the transport and recombination of band edge carriers. These findings illustrate a fundamental difference between HOIPs and conventional photovoltaic semiconductors and demonstrate the importance of understanding lattice dynamics in the effort to develop metal halide perovskite optoelectronic devices.

## Introduction

One of the distinguishing properties of hybrid organic-inorganic perovskites (HOIPs) compared to conventional semiconductors is the presence of considerable dynamical disorder in the lattice (1–3). In methylammonium lead iodide ($MAPbI_3$, herein MAPI; MA = methylammonium = $CH_3NH_3^+$), the organic cation undergoes rapid reorientations about its average lattice position (4, 5). Complicating the dynamical picture, these cation motions couple strongly to the inorganic lead-halide inorganic framework, as observed in both theory (6) and experiments (7). The lattice dynamics are relevant to both the cooling and transport of photoexcited charge carriers because these phenomena are influenced by the coupling between electronic excitations and lattice vibrations, i.e. electron-phonon interactions (8, 9).

The lattice dynamics in MAPI have attracted significant research attention. Most studies have focused on the dynamics of the MA cation and the related tetragonal-to-cubic perovskite phase transition at 327K (10), with recent work suggesting the existence of local dynamic tetragonal domains above the phase transition temperature (11–14). The measured phonon dispersions in MAPI agree well with first-principles lattice dynamics calculations (13–16). Phonon lifetimes are critically important as well, as they provide insight into both electron-phonon and phonon-phonon scattering (17). Recent preliminary reports suggest that the phonon lifetimes in MAPI are significantly shorter than those in conventional semiconductors. Theoretical studies employing molecular dynamics (18) and phonon many-body perturbation theory (19) predict short lifetimes in MAPI due to anharmonic phonon-phonon scattering. Recent experimental studies support these findings, but the reported lifetimes span 1-2 orders of magnitude (14, 16), with acoustic phonon lifetimes ranging from 1-20ps (14). The paucity of precise measurements limits current understanding of charge carrier cooling and transport, both of which are fundamental aspects of the optoelectronic properties of HOIPs.

Phonon lifetimes are directly related to the lattice thermal conductivity and ultralow thermal conductivity in the range of 0.2-0.5 W $K^{-1}$ $m^{-1}$ has been measured for MAPI in both single crystals and polycrystalline samples (20, 21), with low values also computed from molecular dynamics (MD) simulations (18, 22, 23) and third-order perturbation theory (19). The thermal conductivity of MAPI is 100-500× smaller than that of conventional semiconductors such as GaAs (37 W $K^{-1}$ $m^{-1}$) and Si (124 W $K^{-1}$ $m^{-1}$) (24), and resembles that of amorphous polymers (25). While low thermal conductivity makes HOIPs exciting candidates for thermoelectric applications (22), inefficient thermal transport also has important consequences for the operation of optoelectronic devices. If phonons are slow to dissipate heat, this can create nonequilibrium phonon populations and affect carrier relaxation and scattering processes. If the relaxation dynamics are controlled, these mechanisms could potentially enable hot carrier solar cells capable of exceeding the Shockley-Queisser limit (26–28).

A detailed understanding of phonon-phonon interactions and thermal conductivity in MAPI is limited by the lack of precise measurements of the phonon lifetimes and, more generally, of the lattice dynamics. We have measured the acoustic phonon dispersions and lifetimes in MAPI along two branches, in two Brillouin zones, and in the orthorhombic, tetragonal, and cubic phases of MAPI using high-resolution neutron spectroscopy. The use of large, fully deuterated single crystals of $(CD_3ND_3)PbI_3$ ($d^6$-MAPI) facilitated our study by reducing the strong incoherent scattering from hydrogen atoms. Our data reveal extremely short acoustic phonon lifetimes of 20 ps close to the Brillouin zone center and < 1 ps near the zone boundary. First-principles modelling of the lattice dynamics with third-order perturbation theory indicates that three-phonon scattering processes are responsible for these short lifetimes and shows that these interactions are facilitated by strong phonon-phonon interactions and a high density of low-energy optical phonon modes. Furthermore, we fit our measured phonon dispersion relations to extract the momentum-resolved acoustic phonon group velocity and determine the acoustic phonon mean free paths, which range from 0.5-7 nm, in some cases smaller than a single unit cell (1). These short lifetimes confirm the inability of acoustic phonons to dissipate heat efficiently in MAPI. Finally, our measurements of the quasielastic scattering in deuterated $d^6$-MAPI shows reorientational cation dynamics with 1 ps lifetimes in the cubic and tetragonal phases, which are similar to those observed in protonated samples (4, 5). This suggests that the lattice dynamics we observe are consistent with those of non-deuterated MAPI.

# Results and Discussion

In our experimental geometry, we vary the energy transfer $E_T$ while keeping the momentum transfer **Q** constant, i.e.: we measure constant-**Q** scans. Each constant-**Q** scan captures the interaction between incident neutrons and the crystal lattice as a function of energy transfer. Positive (negative) $E_T$ describes neutrons losing energy to (gaining energy from) the lattice. Each constant-**Q** scan exhibits a central peak at $E_T$=0 containing both an elastic and quasielastic scattering (QES) component. In addition, if a phonon mode with energy $E_{ph}$ has a sufficient structure factor S(**Q**, $E_{ph}$), the constant-**Q** scan will exhibit peaks centered on positive and negative $E_{ph}$ corresponding to phonon creation and annihilation respectively. Figure 1a illustrates how constant-**Q** scans at different momentum transfer enable us to map phonon modes through the Brillouin zone.

Fitting the constant-**Q** scans to an appropriate model enables us to extract the phonon energies and linewidths as a function of reduced wavevector *k*, where *k* represents the distance from **Q** to the Brillouin zone center measured in reciprocal lattice units (r.l.u.) (29). For the central peak, the elastic incoherent component corresponds to static diffuse scattering and is well described by a Gaussian lineshape with a fixed width determined by the instrumental resolution. The QES component corresponds to non-propagating (or incoherent) lattice dynamics and is fit to a Lorentzian profile. The phonon peaks are well described by Voigt profiles, where the Lorentzian and Gaussian components correspond to the intrinsic phonon linewidth and instrument resolution, respectively. The amplitudes of the phonon creation and annihilation peaks are related by detailed balance (29). Figure 1b illustrates the fit for a representative constant-**Q** scan. Details of the fitting procedure are provided in the Methods and Supplementary Information appendix (SI).

Whitfield et al. have reported that partial and full deuteration do not affect the crystal structure of MAPI nor the phase transition temperatures (12). We measured the elastic neutron scattering from our deuterated crystal and observed the orthorhombic-tetragonal transition at 163 ±2 K and the tetragonal-cubic transition at 327.5 ±1 K (Figure S1), values identical to those in protonated MAPI (1). This suggests similar dynamic behavior of deuterated and protonated MAPI, which is further supported by our measurements of the quasielastic scattering (see below).

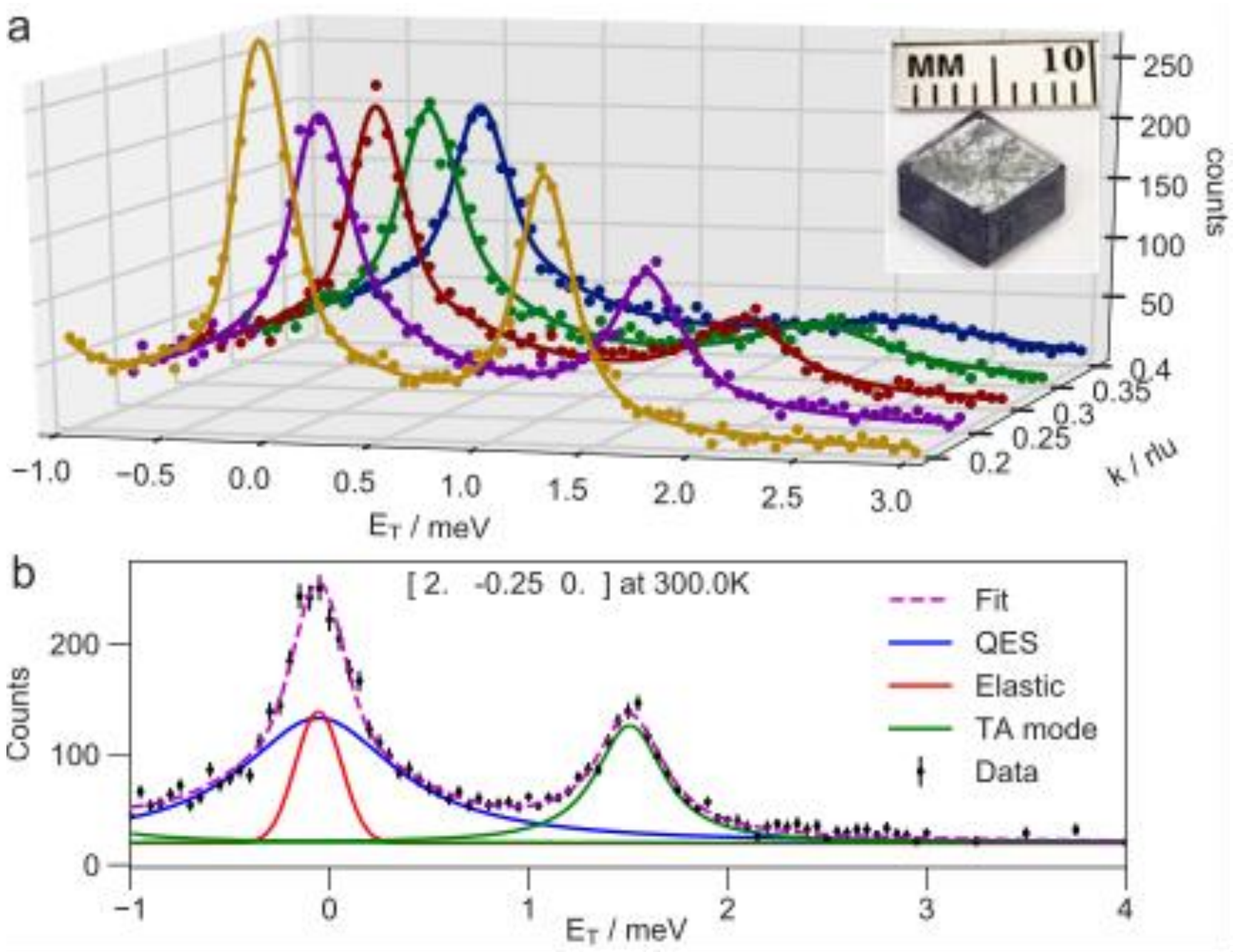

Figure 1: Constant-**Q** scans and fits. a) Constant-**Q** scans and their fits measured along the [2, -k, 0] direction. As k increases, the TA phonon energy increases and the linewidth broadens. The curves are colored yellow (k=0.2), purple (0.25), red (0.3), green (0.35), and blue (0.4). Inset: fully deuterated $d^6$-MAPI single crystal weighing 420 mg. b) Constant-**Q** scan at [2, -0.25, 0] (purple curve in a) with fit and components shown.

The transverse acoustic (TA) phonon dispersion along [110] ($\Gamma - M$) in the cubic phase (350K) and the TA dispersion along [100] ($\Gamma - X$) in the cubic, tetragonal (300K) and orthorhombic (140K) phases are shown in Figure 2. The momentum transfer **Q** in all three phases is referenced relative to the pseudo-cubic unit cell for simplicity (see SI). Both phonon branches soften slightly (i.e. shift to lower energy) on heating from the orthorhombic to the tetragonal phase, but they show little difference between the tetragonal and cubic phases. Constant-**Q** scans measured in the orthorhombic phase show splitting of the TA phonon, likely due to crystal twinning (see SI text and Figure S2). In Figure 2, we plot the lower energy of these TA peaks, so this plot represents a lower bound on the TA mode energies in the orthorhombic phase and a lower bound to the softening that occurs at the phase transition. Our observation of equivalent phonon dispersions in the tetragonal and cubic phases is consistent with studies that report the microscopic equivalence of these phases around the phase transition temperature (11–14). For example, Beecher *et al.* propose that MAPI is made up of small dynamic tetragonal domains at 350K, even though the average structure is cubic (14). Our measurements qualitatively support this picture.

We have also measured the longitudinal acoustic (LA) phonon dispersion along [100] ($\Gamma - X$). However, the spectrometer vertical resolution is sufficiently relaxed that we observe contamination from the TA mode that complicates the fitting (30). We include the LA phonon in the dispersions plotted in Figure S3 and other measurements of the LA phonon are shown in subsequent plots of the SI.

We compare our measurements to the phonon dispersion of the orthorhombic structure calculated using *ab initio* lattice dynamics based on density functional theory (see Methods) (19). The calculated TA dispersions along $\Gamma - X$ and $\Gamma - M$ are in excellent quantitative agreement with the experimental measurements, and the calculations also give a good reproduction of the LA dispersion measured along $\Gamma - X$ (see Figure S3). While identical calculations have been performed on the cubic phase (15, 19), these may not be as representative of the actual structure: first, soft phonons have been experimentally identified in the cubic phase (13, 14), which lead to imaginary harmonic modes in the calculations (6); and second, the harmonic phonon calculations are based on a static structure that does not account for the cation reorientational dynamics in the cubic and tetragonal phases (4, 5). In contrast, the orthorhombic structure has no imaginary modes (19) and a well-defined cation orientation (10), so we expect calculations for this phase to give a good reproduction of the structural dynamics. Furthermore, as described above, the high-temperature cubic phase exhibits local domains of lower symmetry, so the orthorhombic structure may be a better theoretical model to describe lattice dynamics on this basis. The complete calculated phonon dispersions are plotted in Figure S4.

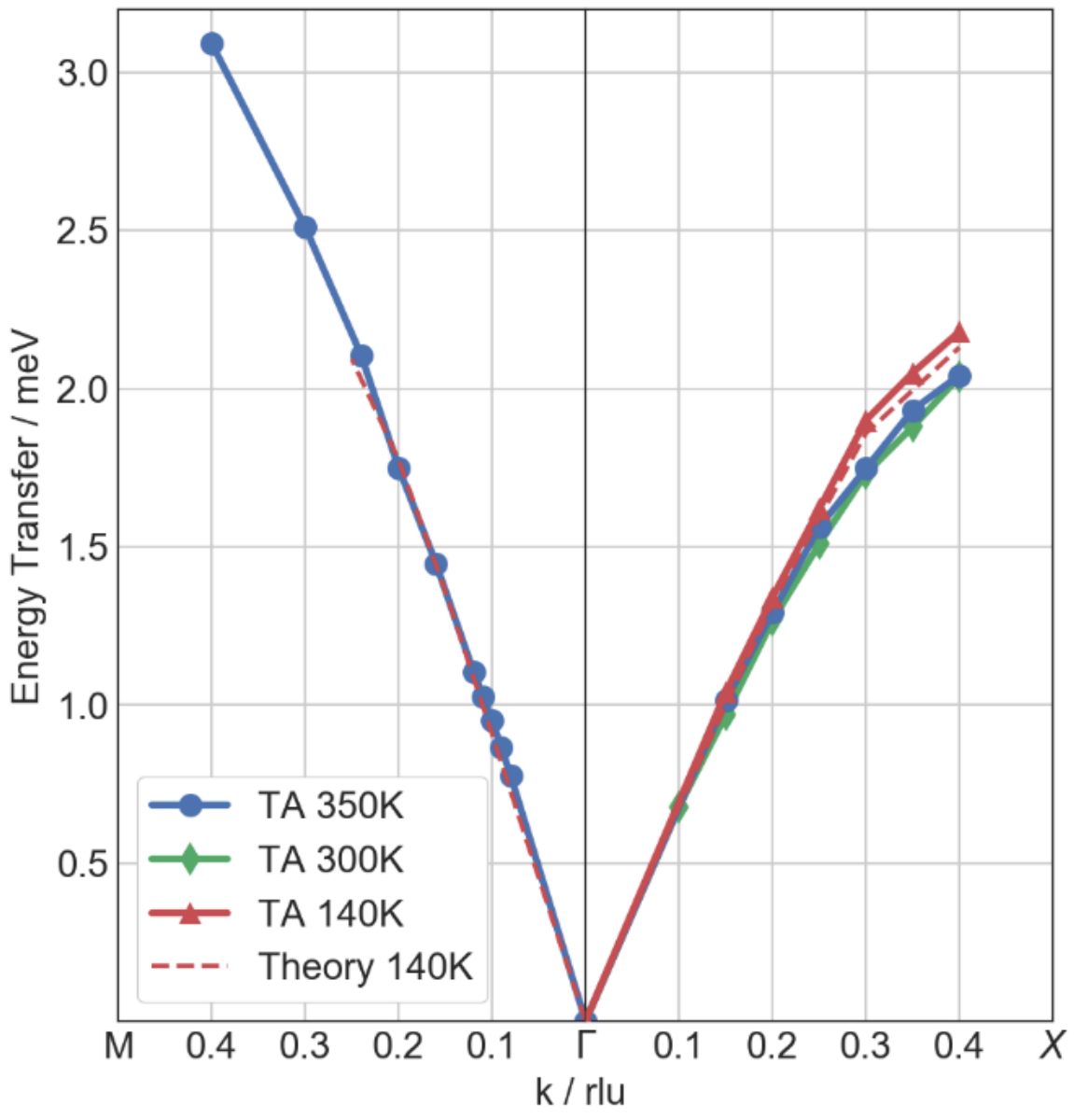

Figure 2: TA phonon dispersions along $M - \Gamma - X$. Blue circles, green diamonds, and red triangles correspond to the cubic (350 K), tetragonal (300 K) and orthorhombic (140 K) phases, respectively. The dashed lines show the dispersion simulated from *ab initio* lattice-dynamics calculations.

Our neutron scattering measurements can also provide insight into the contribution of specific phonon modes to the thermal conductivity *via* the phonon lifetimes. From a theoretical standpoint, the thermal conductivity is the product of heat capacity, phonon group velocity, and phonon mean free path, summed over phonon branches and averaged over wavevectors (31). Using our harmonic lattice dynamics calculations and calculated phonon lifetimes (see Methods), we calculate the room temperature lattice thermal conductivity of MAPI to be 0.086 W $m^{-1}$ $K^{-1}$, in rough agreement with experiments and MD simulations. In Figure 3, we plot the phonon density of states (DoS) overlaid with the cumulative lattice thermal conductivity to show the contributions from modes in different energy ranges. The cumulative lattice thermal conductivity, i.e. the thermal conductivity corresponding to phonon modes with energy ≤ E, demonstrates that acoustic phonon modes in MAPI (E < 3 meV, see Figure S4) contribute relatively little to the thermal conductivity, with the majority of heat transport occurring via low-energy optical modes (E < 10 meV). This is distinct from conventional semiconductors such as GaAs and CdTe, in which acoustic phonons are responsible for conducting the majority of the heat (19). Figure 3 also reveals substantial coupling between phonon modes corresponding to the inorganic cage (Pb-I octahedra) and the organic molecule, as evidenced by the overlap between the partial DoS projected onto the cage and MA cation atoms between 8-14 meV. This overlap suggests a large cross section for energy-conserving phonon-phonon scattering events, which we discuss below.

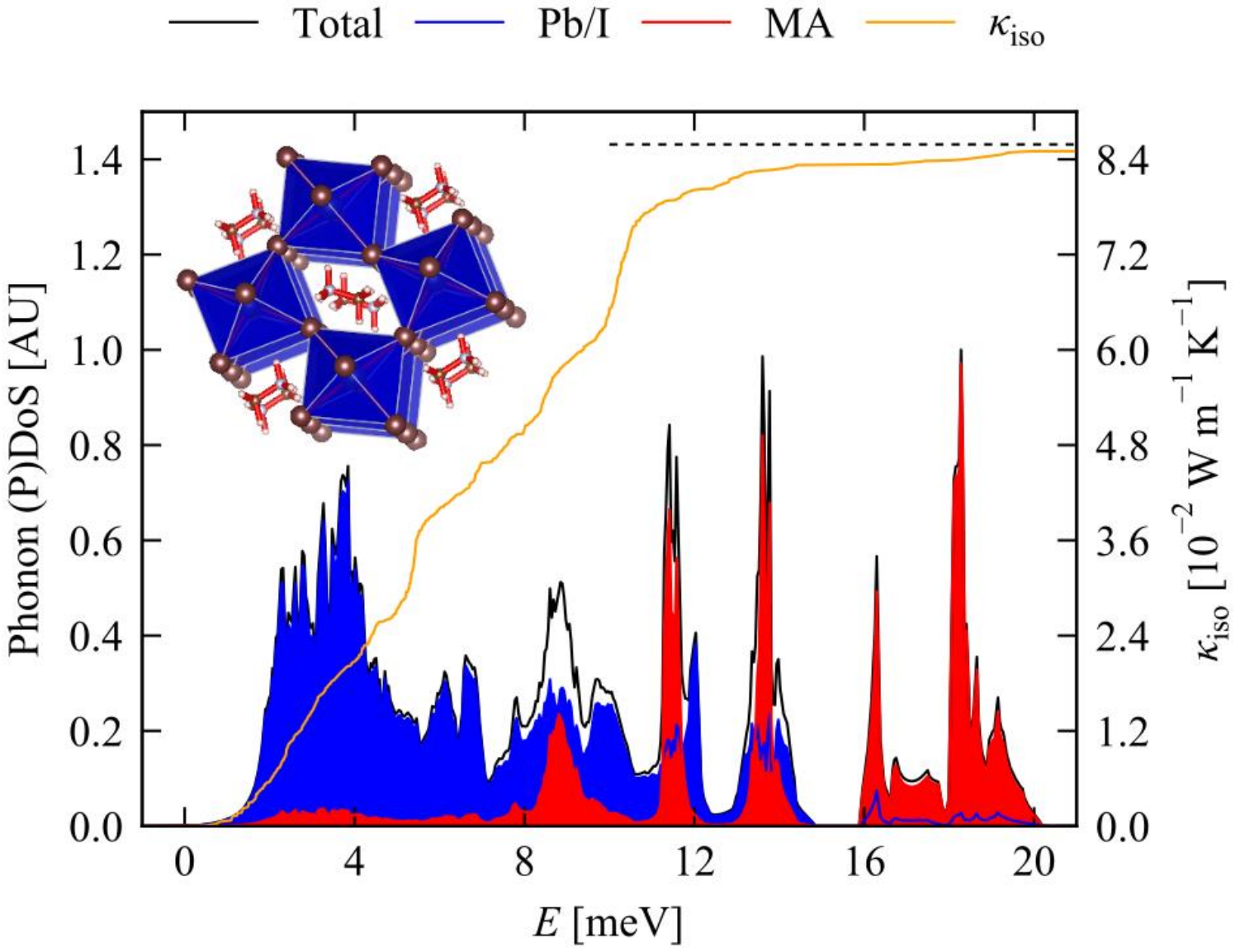

Figure 3: Phonon density of states of MAPI for phonon energies from 0-20 meV (black) with projections onto the Pb-I cage (blue) and MA cation (red). The cumulative lattice thermal conductivity $\kappa_{iso}$ at 300 K is overlaid in orange. The dashed black line shows the total thermal conductivity summed over all modes, including the “pure” molecular modes at higher energies. Inset: schematic of orthorhombic-phase MAPI crystal structure with Pb-I octahedral cage in blue and MA cation in red.

As shown in Figure 3, our calculations suggest that acoustic phonons in MAPI transport heat inefficiently. In order to verify this, we experimentally investigated the contribution of the measured TA phonon modes to the lattice thermal conductivity by considering the phonon group velocity and mean free path, whose product is related to the thermal conductivity as described above. The slope of the dispersion curves in Figure 2, $d\omega/dq$, gives the group velocity, which is the speed at which a phonon mode can propagate energy through the lattice. To calculate the momentum-resolved acoustic phonon group velocity from our measurements, we fit a 2nd order polynomial to the measured dispersion relation and then evaluated the derivative as a function of *k*. At *k*=0, These fits reveal zone center group velocities (speed of sound) ranging from ~2400 m/s for the LA mode to 1200 m/s for the TA mode along $\Gamma - X$. This is in quantitative agreement with measurements on the LA mode from pump-probe spectroscopy measurements (32). These group velocities are somewhat slow: in comparison, Si has a speed of sound ranging from 4500-9000 m/s (33).

The phonon mean free path is the product of the momentum-resolved phonon group velocity and phonon lifetime; the latter describes the average lifetime of a phonon mode before scattering. Phonon lifetimes are related to the phonon energy linewidth via the Heisenberg uncertainty relation, as phonon-defect scattering, phonon-electron scattering, and the anharmonic processes responsible for phonon-phonon scattering all serve to broaden the phonon peaks measured in constant-**Q** scans (34). After accounting for instrumental broadening, the intrinsic phonon half width at half maximum (HWHM) gives the reciprocal of the phonon lifetime (35, 36). Measuring phonon lifetimes is challenging due to the strict requirements on instrumental energy resolution. While Raman scattering techniques can achieve resolution in the μeV range, and are thus capable of resolving lifetimes as long as nanoseconds, these measurements only probe zone-center optical modes (16). Inelastic X-ray scattering techniques can probe phonon modes throughout the Brillouin zone, but they are limited to energy resolutions of 1-2 meV (37), which are insufficient to resolve picosecond lifetimes.

Our high-resolution neutron scattering measurements provide an order-of-magnitude better energy resolution, of order 0.1 meV HWHM, that enables us to measure intrinsic linewidths corresponding to lifetimes shorter than ~20 ps. This technique has been previously employed in superconducting materials to probe phonon lifetimes above the superconducting phase transition (38, 39). In addition, we calculated phonon lifetimes for the orthorhombic perovskite structure using third-order perturbation theory (31). While this approach represents a current "gold standard" for predicting phonon lifetimes from lattice dynamics, we only consider third-order scattering processes, so the calculated lifetimes should overestimate the real lifetimes, which include phonon-phonon scattering to all orders as well as phonon-defect and phonon-electron scattering.

Figure 4 shows our measurements of the intrinsic TA phonon linewidth along $\Gamma - X$ and $\Gamma - M$, together with the instrumental contribution to the phonon peak broadening (see Methods) and our calculations of the TA phonon lifetimes. Our measurements show that the TA phonon broadens dramatically toward the zone boundary, as predicted for anharmonic acoustic phonon decay (40). Throughout the Brillouin zone, the TA phonon lifetimes are extremely short, from 1-20 ps along both directions, and equivalently short for the LA phonon along $\Gamma - X$ (Figure S5). Although first-principles calculations tend to overestimate lifetimes as described above, our calculations show impressive agreement with the measurements, lending a degree of confidence to our theoretical analysis of the thermal transport.

In most of our measurements, the intrinsic linewidths are larger than the instrumental contribution. The exception is the TA phonon along $\Gamma - M$, where lifetimes near the zone center are in the tens of picoseconds, with corresponding linewidths below the resolution limit of our instrument. Error bars represent the standard error from the least-squares fitting algorithm; however, they do not take into account the energy resolution of each data point, and thus will slightly underestimate the standard error. As noted, crystal twinning results in overlapping TA phonon peaks in the orthorhombic phase scans, which complicates the fitting (Figure S2). We report the fitted linewidths of the lower-energy peak and do not include error bars, with the understanding that standard errors may be significantly higher for the orthorhombic lifetimes.

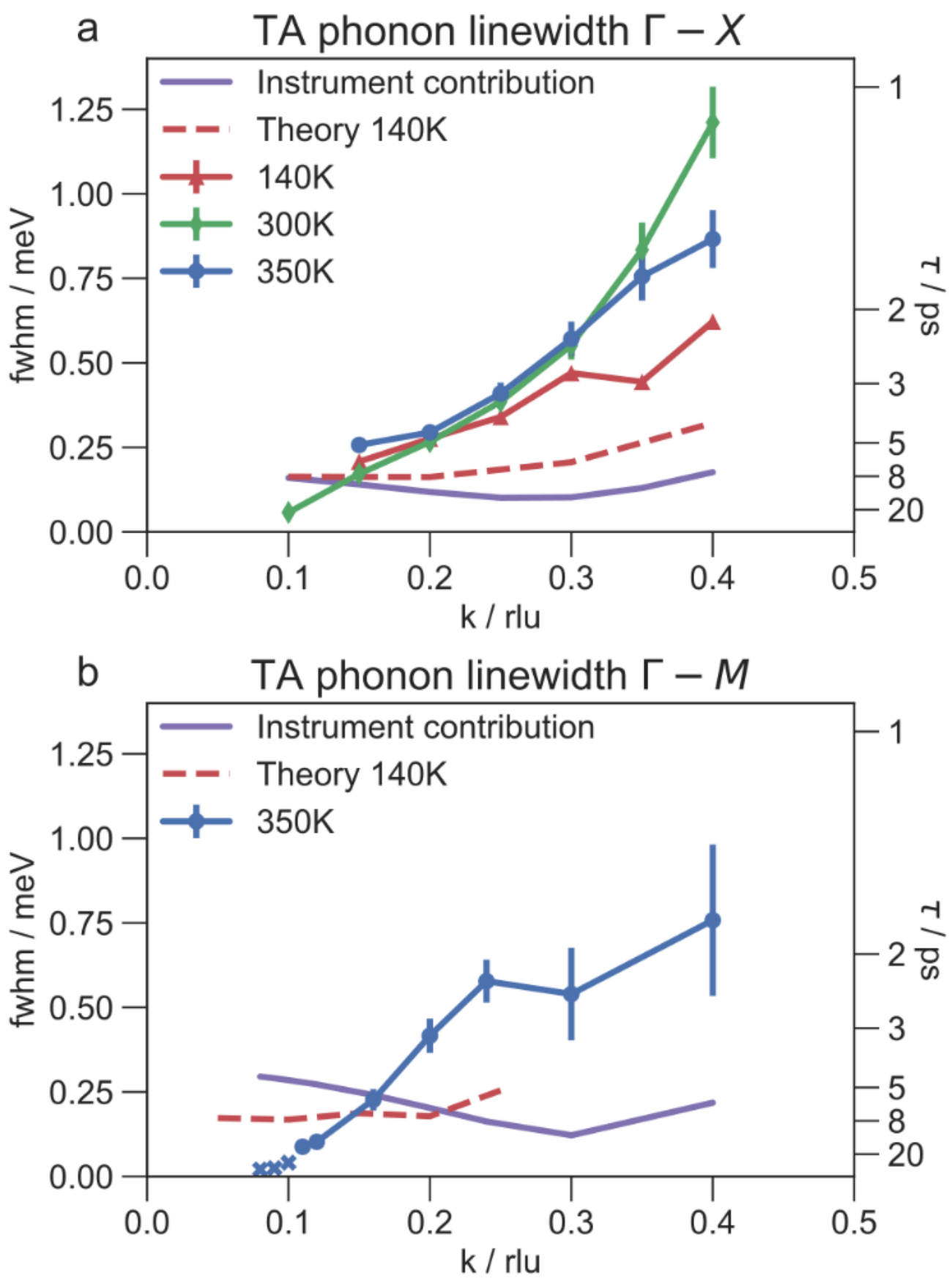

Figure 4: Intrinsic TA phonon linewidths and lifetimes. Error bars represent the standard error in the fitted values, errors in the orthorhombic phase are larger due to crystal twinning. Cross markers denote fitted values below the resolution limit. Dashed lines show the calculated linewidths of the orthorhombic phase. Purple lines show the instrumental contribution to the linewidths. a) TA phonon along $\Gamma - X$ in the orthorhombic (140 K), tetragonal (300 K), and cubic (350 K) phases. b) TA phonon along $\Gamma - M$ in the cubic phase (350 K).

Overall, the TA phonon lifetimes we measure in MAPI are markedly shorter than those in conventional semiconductors. In Si and GaAs, long-lived phonons have correspondingly narrow intrinsic linewidths that are hard to measure given typical resolution limits. However, prior calculations of the acoustic phonon lifetimes in Si and GaAs report acoustic phonon lifetimes of 1 μs near the zone center and 50-100 ps near the zone boundary (41, 42). In comparison, the lifetimes we measure in MAPI are 50-100 times shorter. Additionally, we calculated phonon lifetimes in GaAs and CdTe employing the same methodology we used for MAPI, and we find phonon lifetimes in MAPI to be ~100-500 times shorter than in these conventional semiconductors (Figure S6).

To investigate the microscopic origin of the short acoustic phonon lifetimes in MAPI, we carried out first-principles calculations of the three-phonon scattering processes using the methodology developed in Ref. (31) (see Supplemental Methods). For a given phonon mode, the line broadening can be written as the product of two terms: (1) the two-phonon DoS, which describes the density of energy- and momentum-conserving scattering pathways available for collision (two phonons in, one phonon out) and decay (one phonon in, two phonons out) processes; and (2) an averaged three-phonon interaction strength, which describes the extent of physical coupling to other modes. Figure S6 compares the calculated phonon lifetimes, two-phonon DoS, and three-phonon interaction strength as a function of phonon energy for MAPI, GaAs and CdTe. We find that both the two-phonon DoS and the three-phonon interaction strengths are 1-2 orders of magnitudes higher for acoustic phonons in MAPI than in the other two semiconductors.

The increased density of three-phonon scattering pathways in MAPI is related to the organic cation: the MA cation introduces three translational and three rotational degrees of freedom, which couple to the Pb-I cage and introduce low-energy optical phonon branches in the dispersion, increasing the number of available scattering channels. In contrast, the binary semiconductors CdTe and GaAs have fewer optic branches and more tightly-distributed DoS, resulting in fewer scattering pathways. The origin of the high three-phonon interaction strengths is harder to determine conclusively, but two explanations are likely. Firstly, the soft nature of the inorganic cage leads to large atomic displacements, beyond the harmonic regime where the restoring force is quadratic. Secondly, deformations of the Pb-I cage cause perturbations to the electronic structure, which can couple into the other phonon modes.

To further analyze the line broadening of the acoustic modes, we computed a 2D histogram of the contributions from pairs of phonon modes to the observed line broadening (short lifetimes) of the acoustic and low-energy optical modes in MAPI ($E < 3$ meV, Figure S7). This analysis clearly identifies "hotspots" corresponding to collisions involving bands of modes where motion of the Pb-I cage is coupled to the organic cation, together with substantial broadening from low-energy Pb-I cage modes which we expect to behave anharmonically. This further supports our conclusion that the short phonon lifetimes in MAPI arise from a combination of a high density of scattering pathways related to the MA cation and strong three-phonon interaction strengths due to low-frequency anharmonic motions of the soft Pb-I framework.

Now that we understand the origin of the short lifetimes, we investigate the phonon mean free paths (MFP) by taking the product of the measured phonon group velocity and lifetime. Figure 5 shows our measurement of ultrashort phonon mean free paths from ∼20 nm near the zone center to < 1 nm near the zone boundary. In particular, the MFPs near the zone boundary are shorter than a single unit cell. We find similarly short MFPs for the LA mode (Figure S8). These short MFPs clearly demonstrate the inability of acoustic phonon modes to dissipate heat efficiently in MAPI and reflect the low contribution of these modes to the thermal conductivity evident in the calculations.

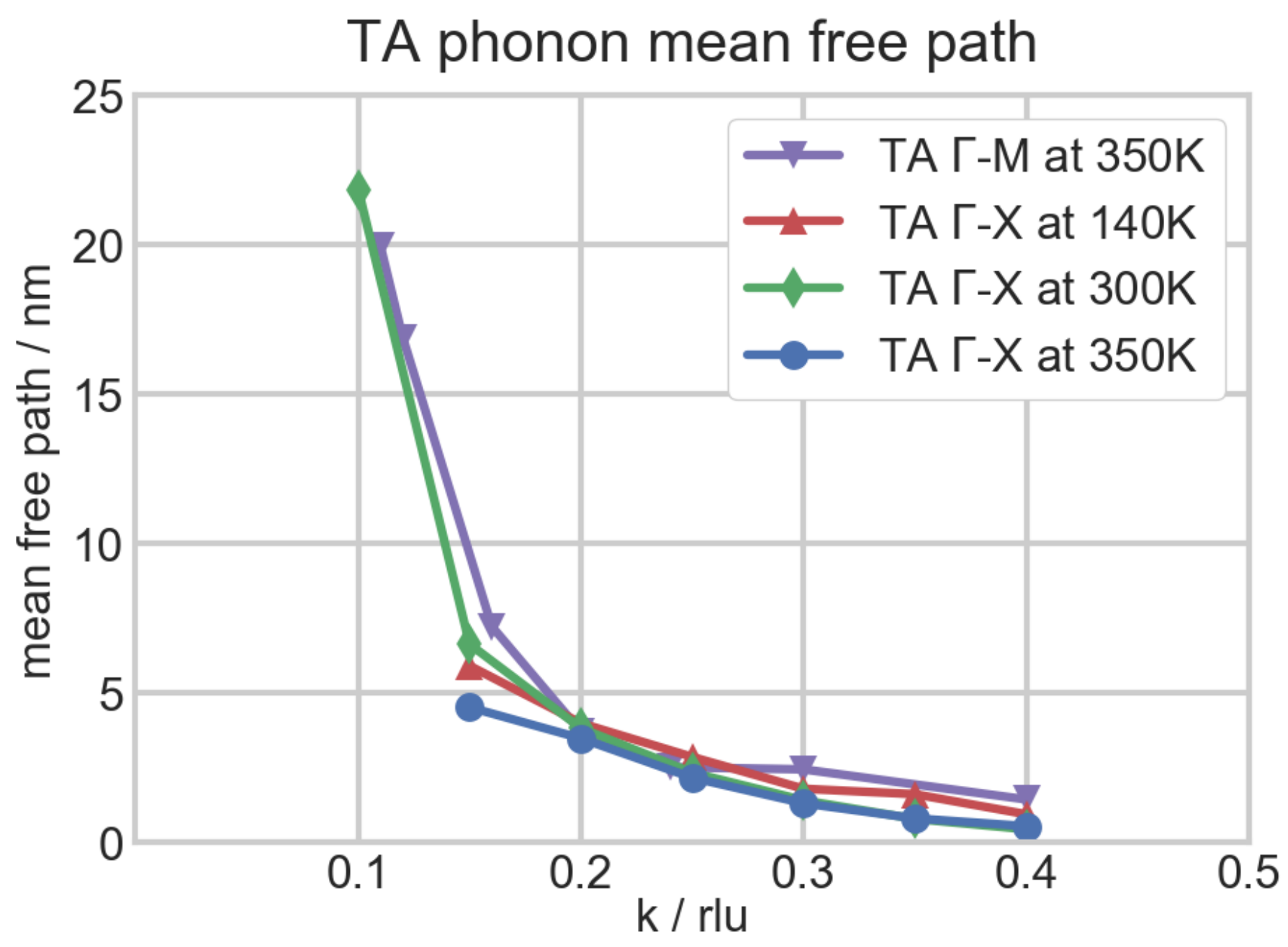

Figure 5: TA phonon mean free paths in MAPI along $\Gamma - X$ and $\Gamma - M$.

On the microscopic scale, the short phonon lifetimes and inefficient thermal transport will have a direct impact on hot carrier cooling (decay of higher-energy carriers to the band edges) and band edge carrier transport. A “hot phonon bottleneck,” i.e. the observation that hot carrier cooling rates decrease dramatically at high carrier concentrations, has been reported in MAPI (26–28). One possible explanation is that inefficient acoustic phonon propagation results in local heating, creating a non-equilibrium population of optical phonons that can effectively reheat carriers (28, 43). Our measurements and calculations support this hypothesis by demonstrating the extreme inefficiency of acoustic phonon heat transport in the MAPI lattice.

The transport of band edge charge carriers also depends on lattice vibrations. A longstanding and important question in the field of HOIPs is how nonradiative recombination rates can be so low in solution-processed HOIPs that likely have a high concentration of defects (2, 44). The microscopic description of trap-assisted Shockley-Read-Hall (SRH) recombination requires the dissipation of an amount of energy equal to roughly half the band gap (~0.8 eV) twice, once during carrier trapping and again during ground-state relaxation. The short acoustic phonon mean free paths and ultralow thermal conductivity in MAPI will limit SRH recombination rates due to local heating. Carrier trapping results in a population of vibrational states localized at the defect center (45), which subsequently decay into acoustic phonons to dissipate the thermal energy to the far field (46). However, in MAPI, these acoustic modes will scatter within picoseconds, leading to persistant local lattice heating within a few nanometers of the trap and facilitating thermal detrapping. Measured nonradiative recombination rates in MAPI suggest lifetimes of hundreds of nanoseconds (44), and thus we speculate that picosecond acoustic phonon lifetimes may play a further role in reducing nonradiative recombination rates *via* carrier detrapping.

Finally, we fit the broad central peak in each constant-**Q** scan to characterize the momentum-resolved quasielastic scattering (QES). QES is sensitive to motions on ps-ns timescales and has been used to probe the cation reorientation in MAPI (4, 5). In the present work, the measured QES intensity is roughly constant throughout the Brillouin zone and dramatically increases in intensity on heating from the orthorhombic phase into the tetragonal phase (Figure S9). These observations suggest that the QES arises from the incoherent scattering associated with cation reorientations. By fitting the QES linewidth, we measure cation reorientation timescales of ~1 ps in the tetragonal and cubic phases, likely corresponding to rotational jumps about the C-N axis. This matches the timescale measured in protonated MAPI (5), suggesting that the deuterated cation does not have a large effect on the lattice dynamics and that our measurements of phonon lifetimes are thus also applicable to protonated MAPI.

## Conclusion

Our measurements show that the acoustic phonon lifetimes in the prototypical hybrid perovskite MAPI are 50-500 times shorter than in conventional semiconductors. We have used high energy resolution neutron scattering to measure the transverse acoustic phonon dispersions as well as the critically important phonon lifetimes, which range from 1-20 ps throughout the Brillouin zone. Anharmonic lattice dynamics calculations reveal that the short lifetimes result from a high-density of low-energy optical modes, arising from the degrees of freedom of the organic cation, together with strong phonon-phonon interactions. In addition to the impact on the thermal conductivity, these short lifetimes likely affect hot carrier cooling and may also play a role in reducing the nonradiative recombination of band edge carriers. Further research into electron-phonon coupling is key to understanding the fundamental limits on optoelectronic properties such as carrier mobilities. As research continues in the field of HOIPs, careful consideration of the lattice dynamics may enable the rational design of new materials in this family for applications in solar cells and other devices.

## Methods

### Preparation of Deuterated (MA-$d_6$)$PbI_3$ Single Crystals

Deuterated ($CD_3ND_3$)I ((MA-$d_6$)I) was prepared as described in the SI. Single crystals of (MA-$d_6$)$PbI_3$ were grown using an adaptation of the method reported by Saidaminov *et al*. (47). A 1:1 molar ratio of (MA-$d_6$)I (0.297 g) and $PbI_2$ (0.830 g) was dissolved in anhydrous $\gamma$-butyrolactone (1.50 mL) at 70 °C. The solution was filtered and a few drops of $D_2O$ were added to aid crystallization and to help maintain the high deuterium content of the solution. The temperature was then slowly increased to *ca*. 95 °C to form seed crystals, and occasionally further increased (up to *ca*. 105°C) to achieve a slow but steady crystallization rate over a maximum of *ca*. 12 h. Crystals were isolated while still hot by transferring into Parabar 10312 oil (Hampton Research), and quickly separating them from the mother liquor. The oil was removed by washing in anhydrous hexanes. These crystals were used to seed larger crystals: the crystal was placed in a fresh precursor solution at 70 °C and the previous procedure was repeated. After 3 such steps, we obtained two black rhombic dodecahedral crystals: a 420 mg crystal with dimensions of *ca*. 8 × 7 × 3 mm and a 698 mg crystal with dimensions of *ca*. 9 x 9 x 4 mm.

### Elastic neutron scattering and inelastic neutron spectroscopy

Elastic scattering data were measured on the thermal neutron triple-axis spectrometer (BT4) at the NIST Center for Neutron Research (NCNR). Additional details are in the SI. Inelastic neutron spectroscopy was performed on the cold neutron triple-axis spectrometer (SPINS) at NCNR. The FWHM incoherent elastic instrumental energy resolution for this configuration was measured using a vanadium standard to be $2\,\Gamma_{resolution} = 0.27$ meV. The Gaussian width of the phonon Voigt function was calculated *via* the Cooper-Nathans formalism (48) and takes into account the slope of the phonon dispersion. Our use of a Voigt function to describe the phonon scattering data therefore represents a one-dimensional convolution of the Gaussian instrumental energy resolution function with the Lorentzian (damped harmonic oscillator) phonon cross section. The Gaussian widths are documented in Table S1. Additional experimental details and a detailed description of the fitting procedure are provided in the SI.

**First-principles calculations of phonon dispersion and linewidths**

Density-functional theory (DFT) calculations were carried out on the orthorhombic phase of MAPI using the VASP code (49) using a similar computational setup to our previous studies (15, 19) (see SI for full details). The starting point for our calculations was the optimized 48-atom unit cell from the study in Ref. (15). The second-order interatomic force constants (IFCs) calculated in a 2×2×2 expansion of the unit cell (384 atoms), which is commensurate with all the symmetry points in the *Pnma* Brillouin zone, were taken from our previous work (19). The third-order IFCs were obtained using the Phono3py code (31) for a single unit cell (i.e. 48 atoms), requiring 10,814 independent calculations, with a displacement step of $2\times10^{-2}$ Å. The harmonic phonon dispersion and atom-projected density of states were obtained by Fourier interpolation, the latter on a uniform $\Gamma$-centered *q*-point grid with 36×36×36 subdivisions. The phonon lifetimes were sampled on a $\Gamma$-centered 6×6×6 grid using Gaussian smearing with a width of 0.1 THz to integrate the Brillouin zone. Further details are provided in the SI.

# Author Contributions

A.G., P.G. and M.T. designed research; I.S. synthesized deuterated crystals with guidance from H.K.; A.G., P.G., I.S. and D.P. performed neutron scattering measurements; A.G. and P.G. analyzed data; J.S., J.F., and A.W. performed calculations; and A.G. wrote the first draft of the paper. All authors contributed to and reviewed subsequent revisions.

# Acknowledgements

The authors thank Hans-Georg Steinrück, Matthew Beard, and Aaron Lindenberg for helpful discussions. A.G. thanks Craig Brown for assistance with planning neutron scattering measurements. A.G. was supported by NSF GRFP (DGE-1147470) and by the Hybrid Perovskite Solar Cell program of the National Center for Photovoltaics funded by the U.S. Department of Energy (DOE), Office of Energy Efficiency and Renewable Energy. Use of the Stanford Synchrotron Radiation Lightsource, SLAC National Accelerator Laboratory, is supported by the U.S. DOE, Office of Science, Office of Basic Energy Sciences under Contract No. DE-AC02-76SF00515. This research was supported by the EPRSRC (grant no. EP/K016288/1), the Royal Society, and the Leverhulme Trust. *Via* our membership of the UK's HEC Materials Chemistry Consortium, which is funded by EPSRC (EP/L000202), this work used the ARCHER UK National Supercomputing Service. We also made use of the Balena HPC facility at the University of Bath, which is maintained by Bath University Computing Services. Work by I.S. and H.K. was funded by a the DOE, Laboratory Directed Research and Development program at SLAC National Accelerator Laboratory (DE-AC02-76SF00515).

# Supplementary Information for

## Acoustic Phonon Lifetimes Limit Thermal Transport in Methylammonium Lead Iodide

Aryeh Gold-Parker[a,b], Peter M. Gehring[c], Jonathan M. Skelton[d], Ian C. Smith[a], Dan Parshall[c], Jarvist M. Frost[e], Hemamala I. Karunadasa[a], Aron Walsh[f,g], Michael F. Toney[b,1]

[a] *Department of Chemistry, Stanford University, Stanford, CA, USA*
[b] *SLAC National Accelerator Laboratory, Stanford Synchrotron Radiation Lightsource, Menlo Park, CA, USA*
[c] *National Institute of Standards and Technology, NIST Center for Neutron Research, Gaithersburg, MD, USA*
[d] *Department of Chemistry, University of Bath, Bath, UK*
[e] *Department of Physics, Kings College London, London, UK*
[f] *Department of Materials, Imperial College London, London, UK*
[g] *Department of Materials Science and Engineering, Yonsei University, Seoul, Korea*

[1]To whom correspondence should be addressed. Email: mftoney@slac.stanford.edu

## Supplementary Text

### Acoustic mode determination

In any Brillouin zone, we define the reduced wave vector $\boldsymbol{q}$:

$$\boldsymbol{q} = \boldsymbol{Q} - \boldsymbol{G}$$

Where $\boldsymbol{Q}$ is the momentum transfer vector and $\boldsymbol{G}$ is the nearest reciprocal-lattice vector (Brillouin zone center).

The observed neutron scattering intensity from a phonon mode is proportional to the term $(\boldsymbol{Q} \cdot \boldsymbol{u})^2$, where $\boldsymbol{u}$ corresponds to the atomic displacements of that phonon (1). In long-wavelength acoustic modes, all atoms move in phase. For a transverse acoustic (TA) mode, the atomic displacements are perpendicular to the direction of propagation, i.e.: $\boldsymbol{q} \perp \boldsymbol{u}$. In an longitudinal acoustic (LA) mode, the atomic displacements are parallel to the direction of propagation, i.e.: $\boldsymbol{q} \parallel \boldsymbol{u}$.

Due to strong structure factors, we measured all phonon modes near $\boldsymbol{G} = [2,0,0]$. The selection of phonon mode was then determined by the direction of $\boldsymbol{q}$:

The TA branch along $\Gamma - X$ was measured along $[2, -k, 0]$.
The TA branch along $\Gamma - M$ was measured along $[2, -k, -k]$.
The LA branch along $\Gamma - X$ was measured along $[2 + k, 0, 0]$.

### Relating pseudocubic lattice parameters to tetragonal and orthorhombic lattice parameters

All lattice parameters in this manuscript are in pseudocubic lattice parameters for simplicity. In real space lattice parameters, the cubic to tetragonal phase transition corresponds to a doubling of the a3 lattice parameter, a multiplication of a1 and a2 by $\sqrt{2}$, and a rotation of the a1-a2 plane by 45°. In reciprocal space lattice parameters, b3 is halved and b1 and b2 and divided by $\sqrt{2}$, along with the rotation of the b1-b2 plane by 45°.

The lattice parameters are approximately unchanged upon transition from the tetragonal to the orthorhombic phase.

To convert the real space column vector ($a1_{cub}$, $a2_{cub}$, $a3_{cub}$) to the corresponding tetragonal reciprocal lattice coordinates, one can front-multiply the vector by the transformation matrix M:

$$M = \begin{bmatrix} 1 & 1 & 0 \\ -1 & 1 & 0 \\ 0 & 0 & 2 \end{bmatrix}$$

i.e.:

$$M * \begin{bmatrix} a1_{cub} \\ a2_{cub} \\ a3_{cub} \end{bmatrix} = \begin{bmatrix} a1_{tet} \\ a2_{tet} \\ a3_{tet} \end{bmatrix}$$

To convert reciprocal lattice parameters, the inverse of the transformation matrix is used:

$$M^{-1} = \begin{bmatrix} 1/2 & -1/2 & 0 \\ 1/2 & 1/2 & 0 \\ 0 & 0 & 1/2 \end{bmatrix}$$

i.e.:

$$M^{-1} * \begin{bmatrix} b1_{cub} \\ b2_{cub} \\ b3_{cub} \end{bmatrix} = \begin{bmatrix} b1_{tet} \\ b2_{tet} \\ b3_{tet} \end{bmatrix}$$

## Crystal twinning in the orthorhombic phase

After multiple cycles of heating and cooling across the phase transitions, our crystal samples began to show signs of crystal twinning, in which a distribution of domains exists in the low-temperature phases with distinct co-aligned lattice vectors. For example, one domain may have the a1 direction colinear with the a3 direction in another domain. This twinning behavior is common in MAPI and has been observed by other groups performing inelastic scattering measurements (2).

The crystal twinning lead to the appearance of overlapping TA phonon modes in constant-**Q** scans measured in the orthorhombic phase, as shown in Fig. S2. Before beginning a set of constant-**Q** scans, the instrument was aligned using elastic scattering from a nearby Bragg peak. Then, we would move the instrument's focus to a given **Q** where we wanted to measure. However, this reciprocal space alignment would only be correct for one of crystal domains, leading to the measurement of overlapping peaks corresponding to phonon modes at distinct **Q**. Because the twinned TA modes were closely overlapping, there is potentially error in the fitted peak width of the TA modes. For this reason, the actual error in the measurement is likely greater than the standard error of the fit. We explain this in the main text.

## Supplementary Methods

### Preparation of deuterated methylammonium iodide (MA-$d_6$)I.

Partially deuterated methylamine–$d_3$ ($CD_3NH_2$, Sigma Aldrich, product number 486892) was bubbled through stirred methanol in a three-neck flask chilled in an ice bath to create a concentrated solution. An aliquot of the methylamine–$d_3$ solution was then chilled in an ice bath and stirred, then acidified through dropwise addition of hydroiodic acid (57% w/w, stabilized, VWR, product number AAL10410-36). The resulting colorless precipitate was isolated and dried under reduced pressure at 60 °C, yielding crude, partially deuterated $CD_3NH_3I$. The salt was then fully deuterated through recrystallization from methanol-$d_4$ ($CD_3OD$). Solid ($CD_3NH_3$)I was dissolved in anhydrous methanol-$d_4$ and stirred for 3 h at 50 °C. Anhydrous diethyl ether was added and the solution was allowed to cool to room temperature before being placed in an ice bath, yielding colorless crystals. This recrystallization process was repeated 3 times. The final product was dried under reduced pressure at 60°C to yield deuterated (MA-$d_6$)I, which was used to prepare the perovskite single crystals.

### Experimental details of neutron scattering and spectroscopy

The smaller (420 mg) and larger crystals (698 mg) were oriented with the [001] and [01-1] pseudo-cubic axes normal to the scattering plane to provide access to the (HK0) and (HKK) scattering planes, respectively.

Elastic scattering data were measured on the thermal-neutron triple-axis spectrometer (BT4) as a function of temperature at selected values of Q corresponding to the X, M, and R-point zone boundaries. These data, shown in Fig S1 were measured using horizontal beam collimations of 40’-40’-40’-100’ and a fixed incident/final neutron energy of 14.7 meV selected via Bragg diffraction from the (002) reflection of a PG monochromator/analyzer. A highly-oriented pyrolytic graphite (HOPG) filter was placed in the incident beam to eliminate higher order wavelength neutrons.

Inelastic neutron spectroscopy was measured on the cold neutron triple-axis spectrometer (SPINS) with the final neutron energy $E_f$ fixed at 5 meV ($\lambda = 4.045$ Å, $k = 1.553$ Å$^{-1}$) and a liquid-nitrogen-cooled Be filter placed after the sample to remove higher-order wavelength neutrons. The data were measured with horizontal beam collimations of 80’/$k_i$-Open-80’-Open. The (002) reflection of HOPG crystals was used to monochromate and analyze the incident and scattered neutron energies, respectively.

### Fitting constant-Q scans

Constant-Q scans were first reduced by correcting the monitor to account for varying harmonic content of the incident beam, and then normalizing to the incident beam intensity monitor. The fast neutron background was measured at several Q and found to be more than an order of magnitude smaller than that contributed by the sample. To minimize systematic and statistical errors, each constant-Q scan was obtained by summing several independent scans. The resulting error bars represent one standard deviation (±√N, N = total counts). The reciprocals of the errors were used as weights in the fitting procedure.

The reduced data were fit to a model comprised of (a) a resolution-limited Gaussian to describe the elastic incoherent scattering cross section, with FWHM fixed as described in the main paper

Methods; (b) a Lorentzian function to describe the broad quasielastic scattering (QES) cross section, with the FWHM allowed to float; (c) two Voigt functions to describe the acoustic phonon modes for both positive (creation) and negative (annihilation) energy transfers, with intensities related by detailed balance (1), and (d) a flat (hω-independent) background, fit and then fixed for all scans at a single temperature in a given Brillouin zone. All fitting was performed using the lmfit package for Python (3).

The value of $E_T$ corresponding to elastic scattering ("zero point") was allowed to vary within a few hundredths of an meV around 0, accounting for any small potential misalignment of the analyzer ($E_f$) or irreproducibility in the monochromater motors ($E_i$). In each fit, the elastic and QES peaks were centered at the zero point, and the phonon creation and annihilation peaks had their peak centers constrained to be equidistant from the zero point.

For the Voigt profiles fit to phonon peaks, the Gaussian width was fixed to the instrument resolution in Table S1, calculated as described in the main paper Methods. The Lorentzian width of the Voigt profile (the intrinsic phonon linewidth) was then determined by the fit.

## Lattice dynamics calculations

Density-functional theory (DFT) calculations were carried out on the orthorhombic phase of MAPI using the pseudopotential plane-wave formalism as implemented in the Vienna *Ab initio* Simulation Package (VASP) code (4). Electron exchange and correlation was modelled using the PBEsol exchange-correlation functional (5), and a plane-wave basis with a large 700-800 eV cutoff was used with a Monkhorst-Pack *k*-point grid (6) with 3×2×3 subdivisions to model the electronic structure. Projector augmented-wave (PAW) pseudopotentials (7, 8) were used to model the ion cores, with the H 1s, C and N 2s and 2p, I 5s and 5p, and Pb 6s, 6p and 5d electrons being treated explicitly in the valence shell. The precision of the charge-density and augmentation grids was automatically chosen to avoid aliasing errors, and the PAW projection was performed in reciprocal space.

The starting point for our calculations was the optimized crystal structure of the 48-atom unit cell from the study in Ref. (9). Lattice-dynamics calculations were carried out using the finite-displacement method implemented in the Phonopy and Phono3py packages (10–12). The second-order interatomic force constants (IFCs) were calculated in a 2×2×2 expansion of the unit cell (384 atoms), which is commensurate with all the symmetry points in the *Pnma* Brillouin zone, requiring 41 independent calculations (13). The *k*-point mesh was reduced to 2×1×2 subdivisions for these calculations The third-order IFCs were computed in a single unit cell (i.e. 48 atoms), requiring 10,814 independent calculations. Displacement steps of $10^{-2}$ and $2\times10^{-2}$ Å were used to calculate the second- and third-order IFCs, respectively. The harmonic phonon dispersion and atom-projected density of states were obtained by Fourier interpolation, the latter on a uniform Γ-centered q-point grid with 36×36×36 subdivisions. The phonon lifetimes were sampled on a Γ-centered 6×6×6 grid using Gaussian smearing with a width of 0.1 THz to integrate the Brillouin zone.

## Supplementary Figures and Tables

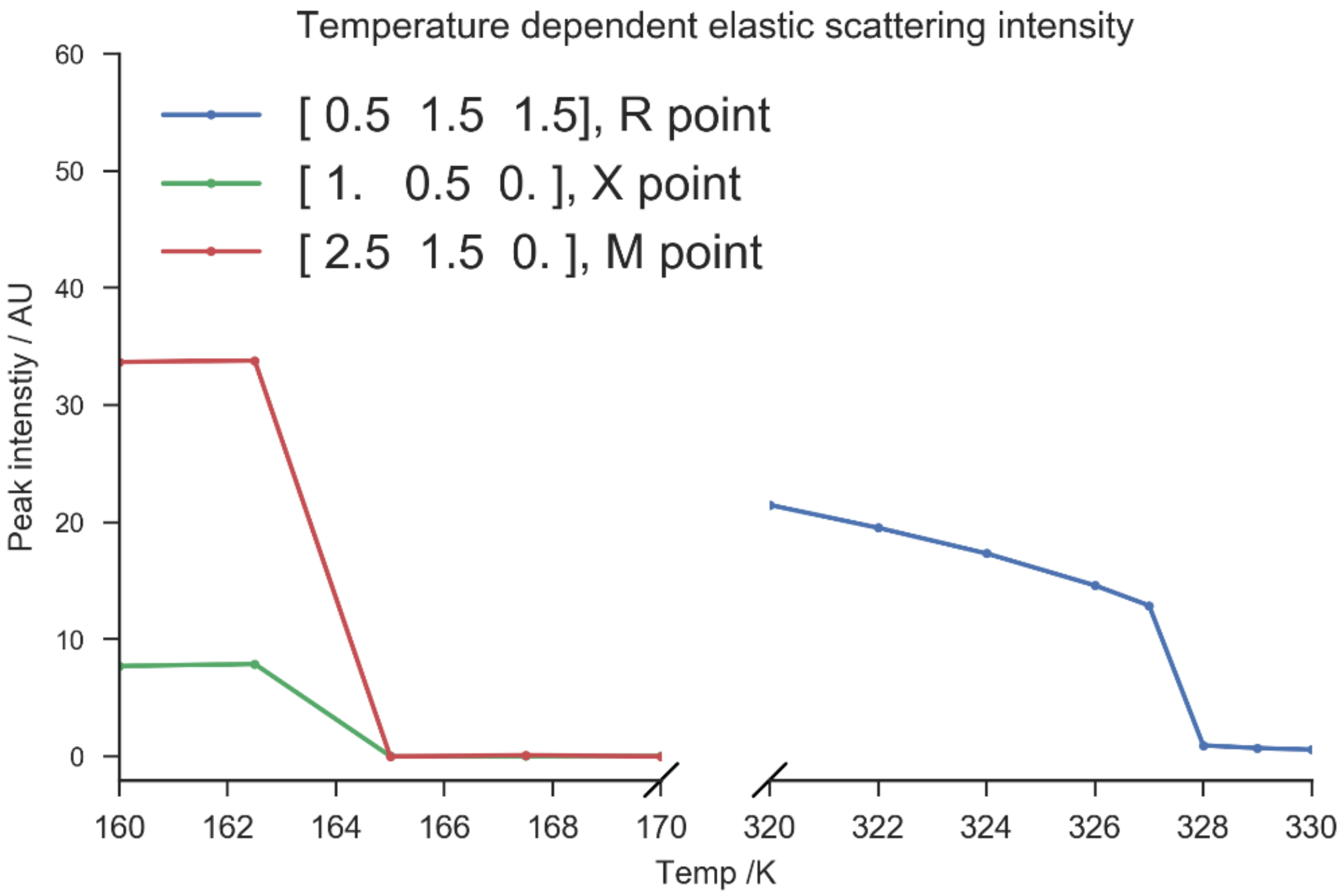

Fig. S1. Temperature-dependent elastic scattering intensity at three high symmetry points (cubic indexing). Rocking curves were taken at each point by varying k in a small range; the resulting peak was fit to a Gaussian and the integrated intensity is plotted. The R point condenses into a Bragg peak at 327 ±1 K, which is the cubic-tetragonal phase transition; the X and M points condense into Bragg peaks at 163 ±2 K at the tetragonal-orthorhombic phase transition. These phase transition temperatures are the same as in protonated MAPI, which are reported to be 327.4 K and 162.2 K (14).

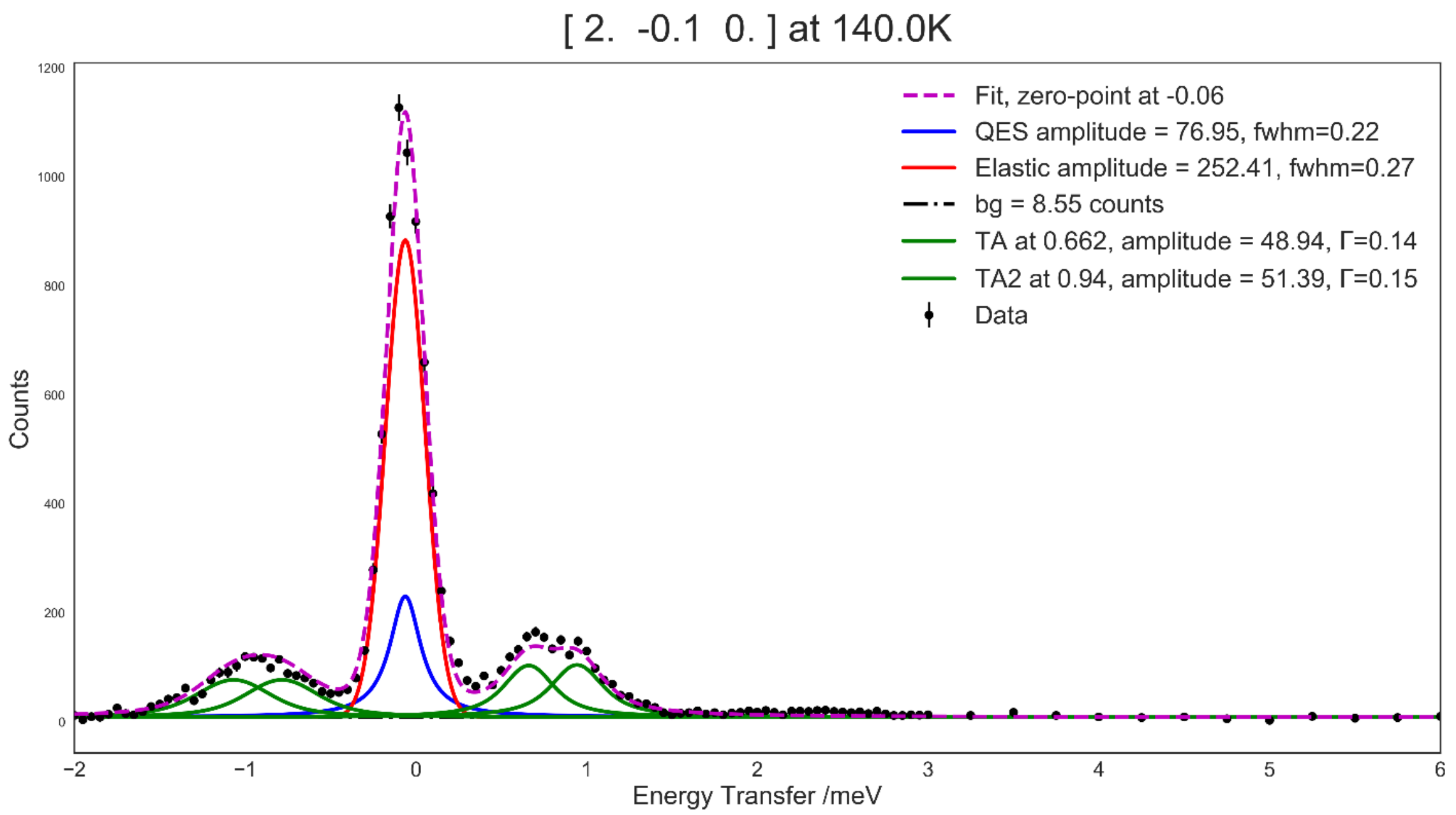

Fig. S2. Example of crystal twinning in the orthorhombic phase. This results in the appearance of overlapping TA modes and complicates fitting of the TA mode in the orthorhombic phase. See supplementary text for more details.

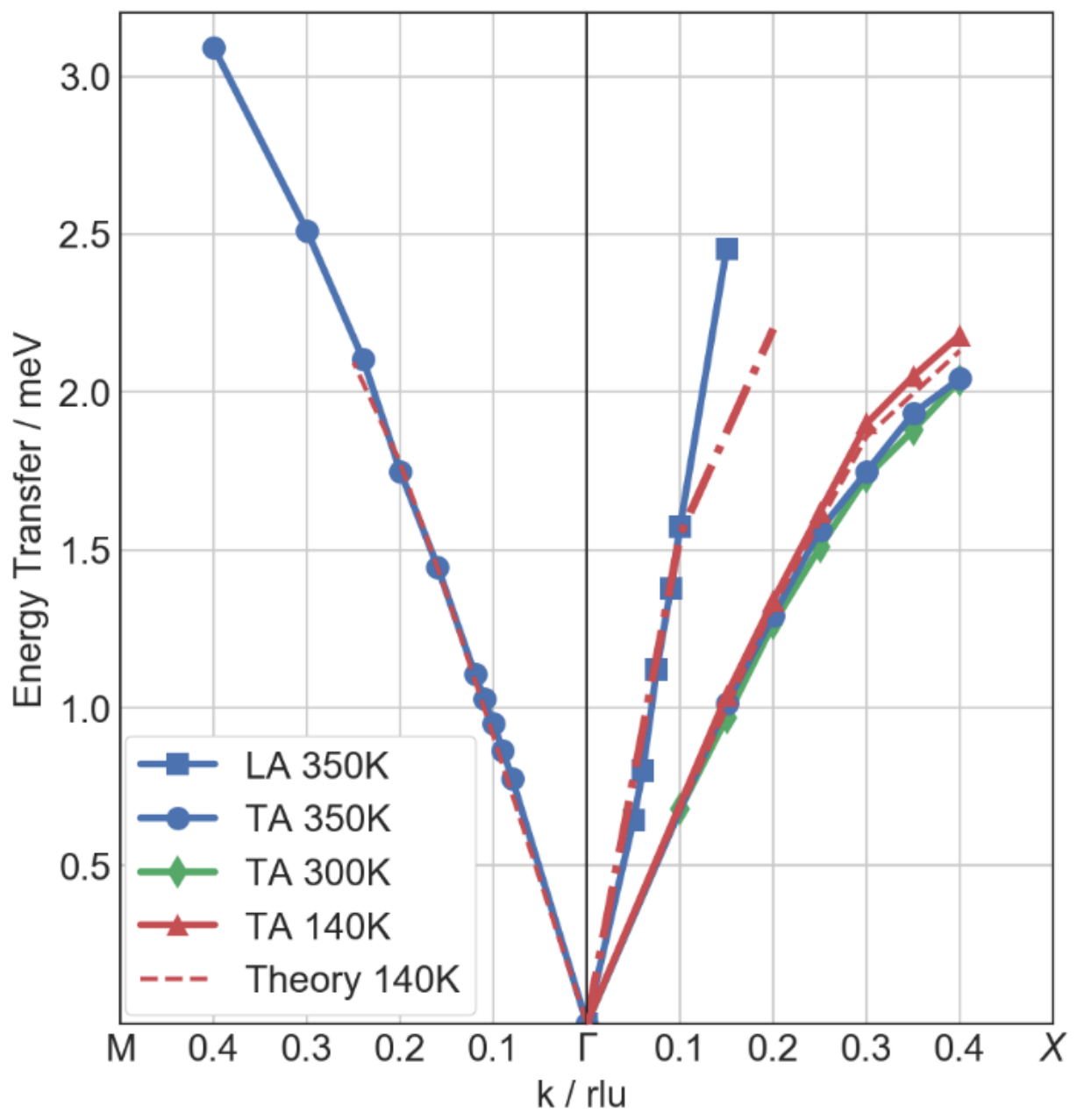

Fig. S3. Acoustic phonon dispersions in MAPI along $\Gamma - X$ and $\Gamma - M$. This is a reproduction of Fig. 2, additionally including the LA mode.

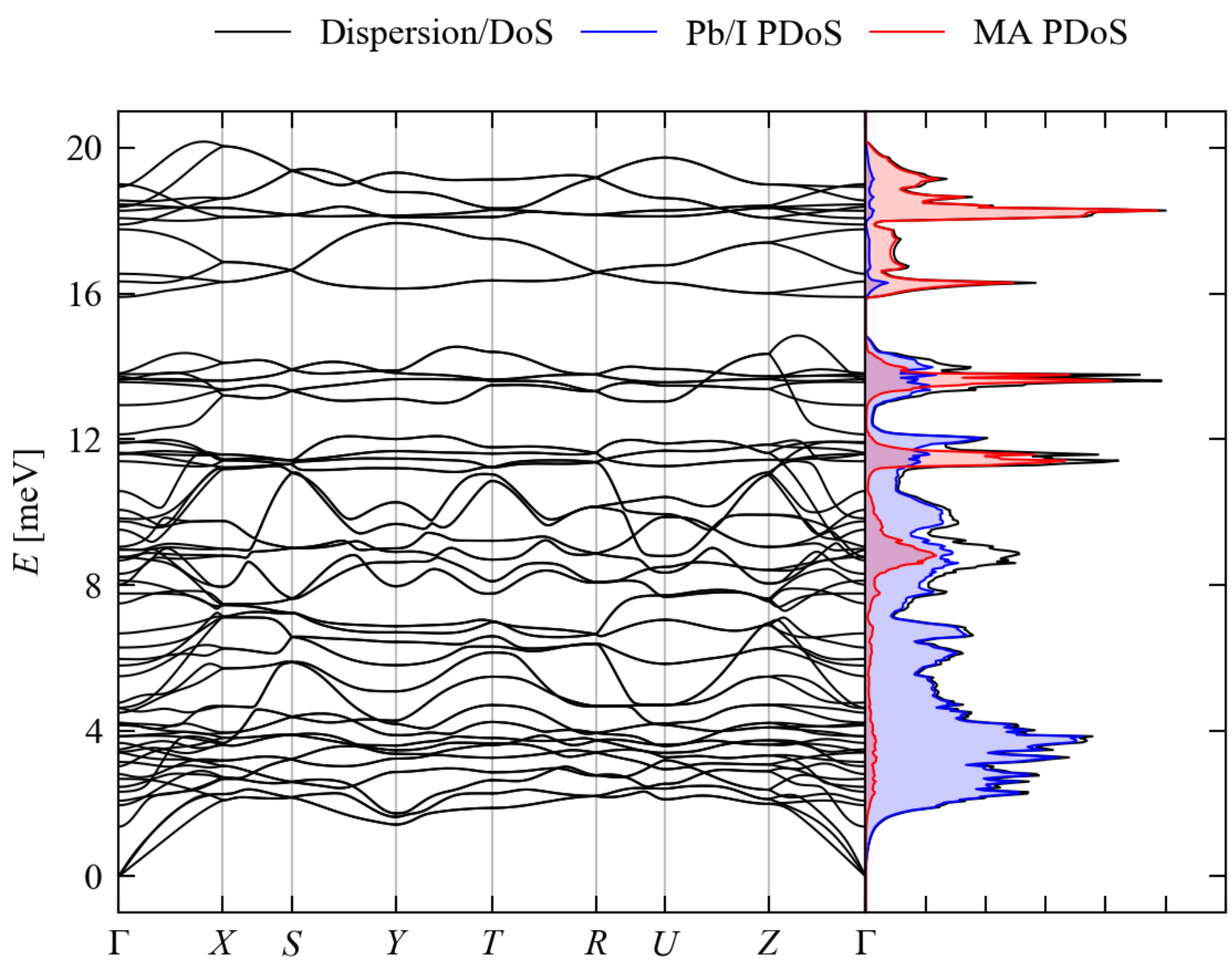

Fig. S4. Simulated phonon dispersion and density of states (DoS; black) for MAPI. As in Fig. 3, the DoS is shown projected onto the PbI cage (blue) and MA cation (red).

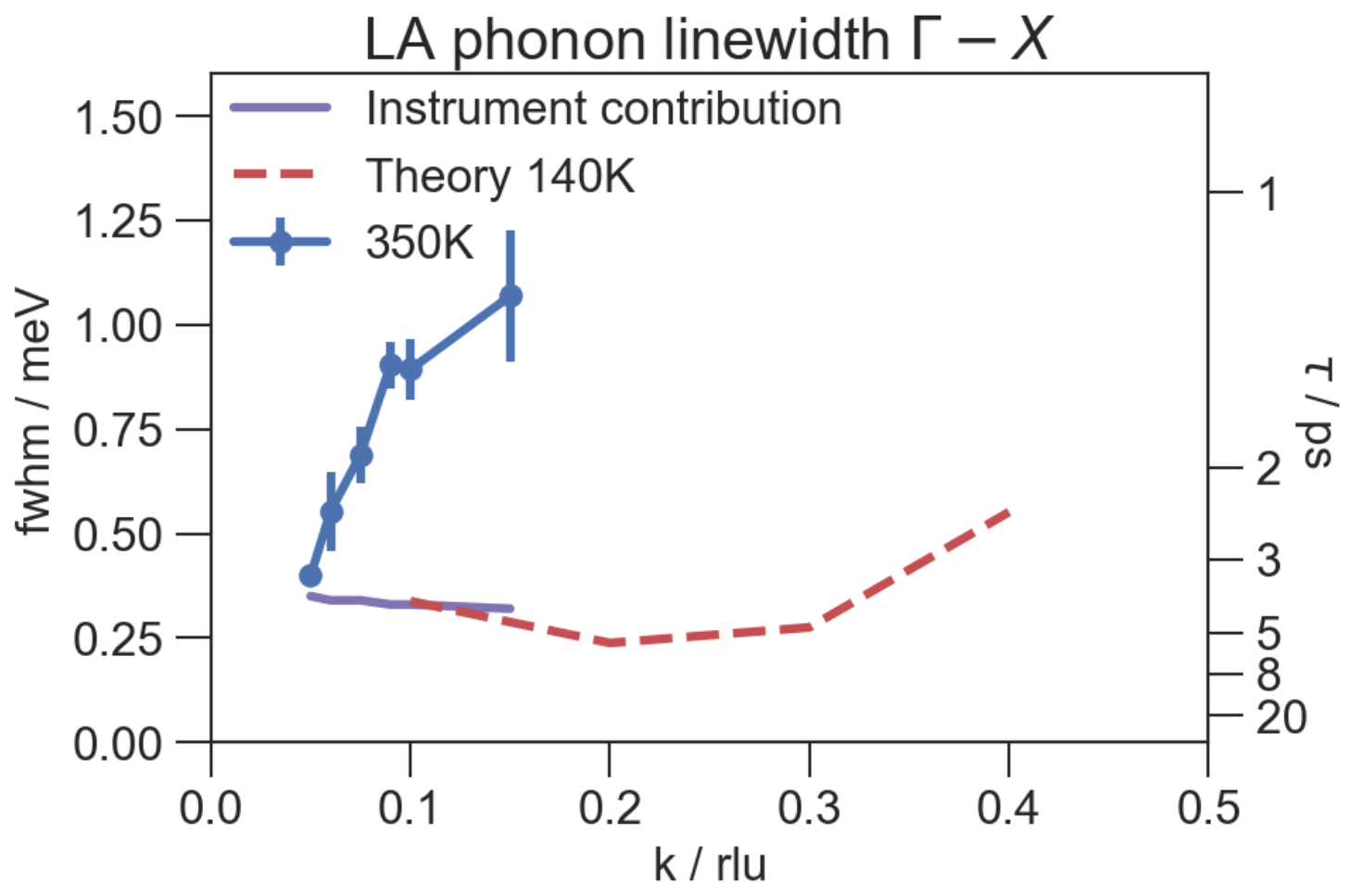

Fig. S5. LA phonon linewidths along $\Gamma - X$ in the cubic phase. Error bars show the standard error in the fitted values. The dashed lines show the linewidths calculated for the orthorhombic phase, while purple lines with no markers show the instrumental contribution to the linewidths. The lack of good agreement between theory and experiment may be due to additional scattering processes not considered in our calculations.

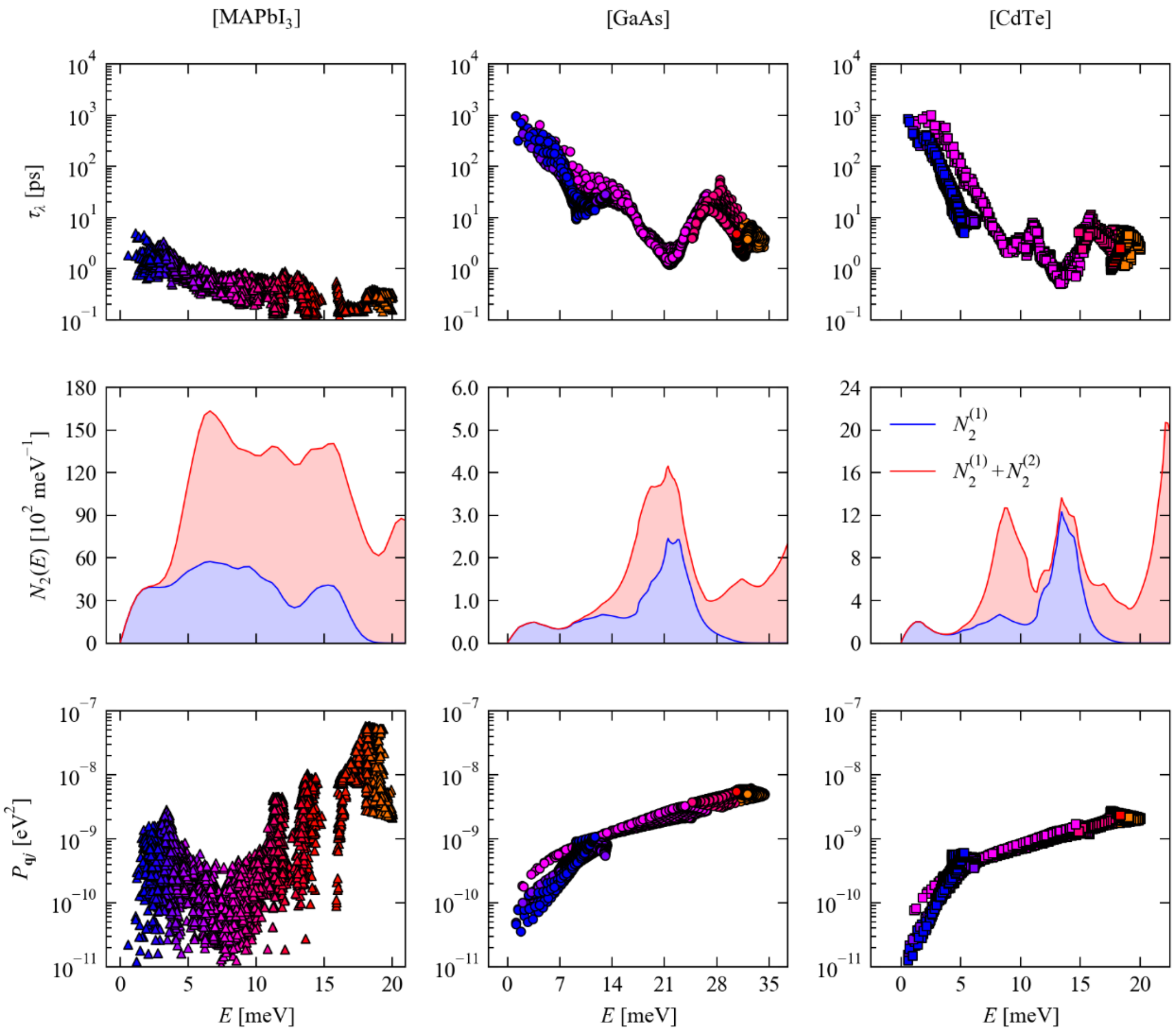

Fig. S6. Comparison of the phonon lifetimes $\tau_\lambda$ to the weighted two-phonon density of states $N_2$, averaged over the Brillouin zone, and the averaged three-phonon interaction strengths $P_{\mathbf{q}j}$ for MAPI, GaAs and CdTe. $N_2^{(1)}$ represents collision pathways (two phonons in, one phonon out) and $N_2^{(2)}$ represents decay pathways (one phonon in, two phonons out). Note the different y-axis scales for the $N_2$ plots This analysis is described in more detail in Ref. (12).

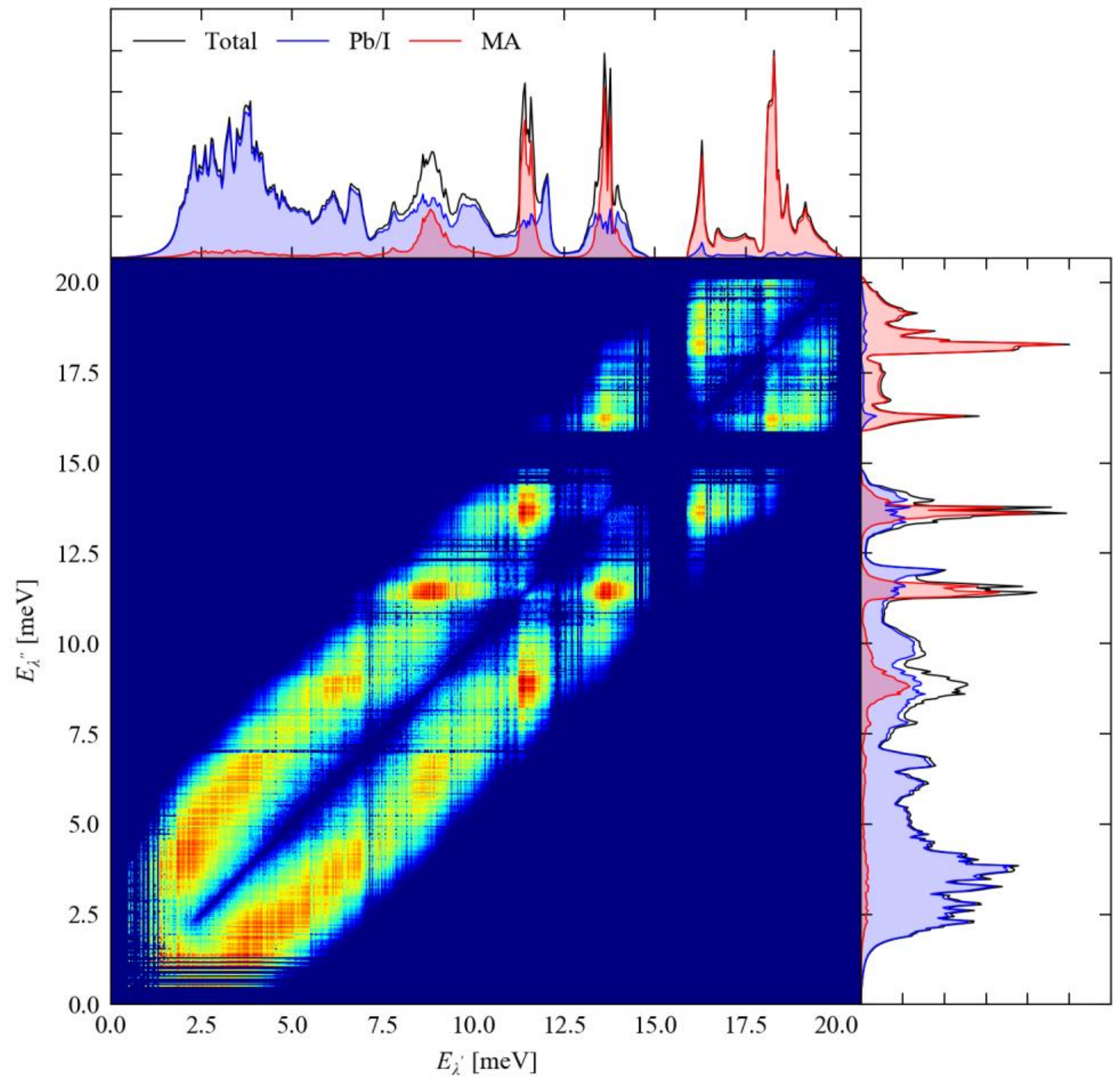

Fig. S7: 2D histogram showing the contributions of pairs of modes with energies $E_\lambda'$ and $E_\lambda''$ to the phonon line broadening of the low-energy modes with $E$ < 3 meV in MAPI, averaged over phonon wavevectors. The color scale is logarithmic and runs from dark blue to dark red, with the latter indicating the largest contributions. For comparison, the phonon density of states (black) and its projection onto the Pb-I cage (blue) and MA cation (red) are shown next to the axes.

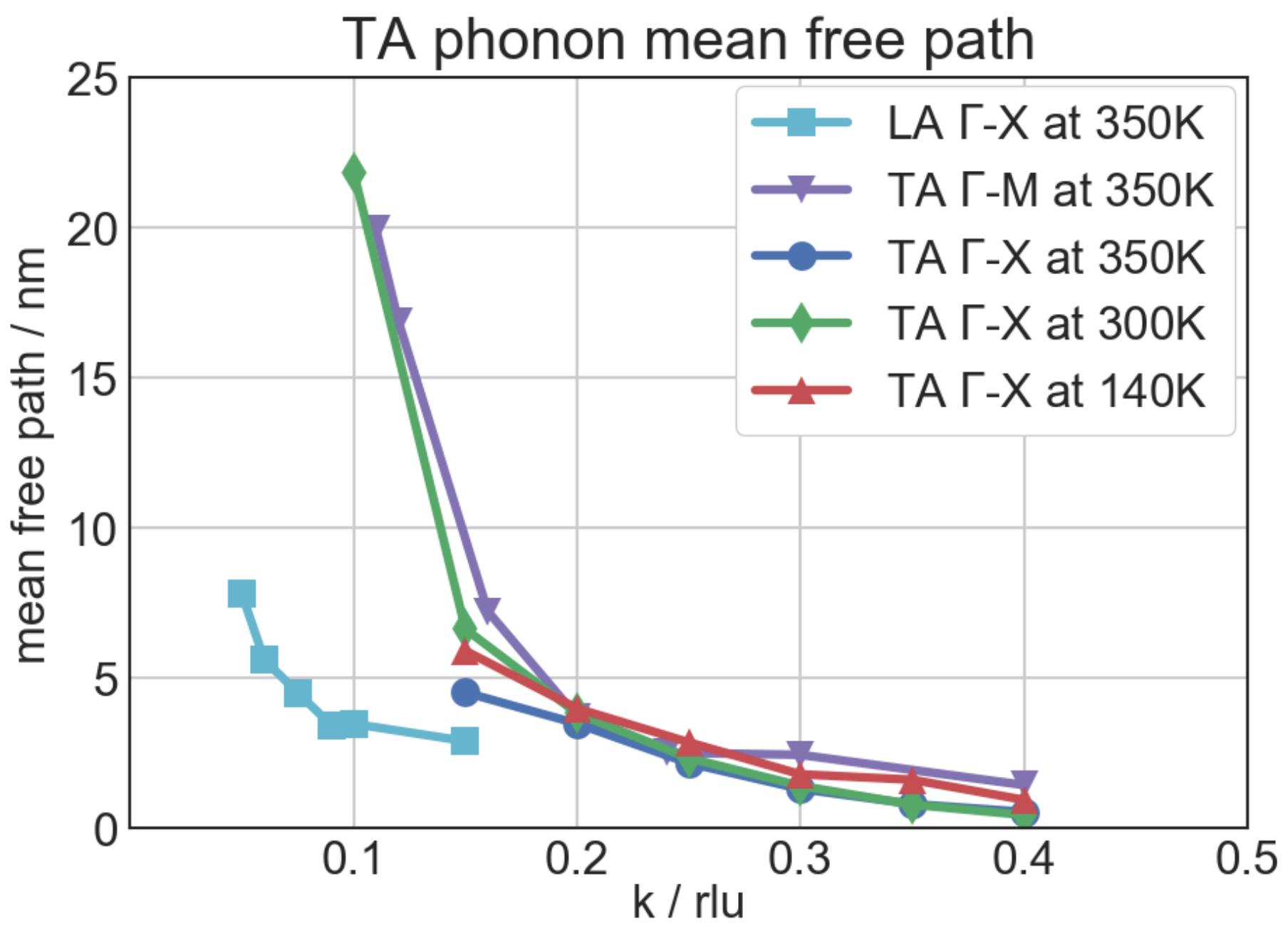

Fig. S8. Acoustic phonon mean free paths in MAPI along $\Gamma - X$ and $\Gamma - M$. This is a reproduction of Fig. 5, additionally including the LA mode.

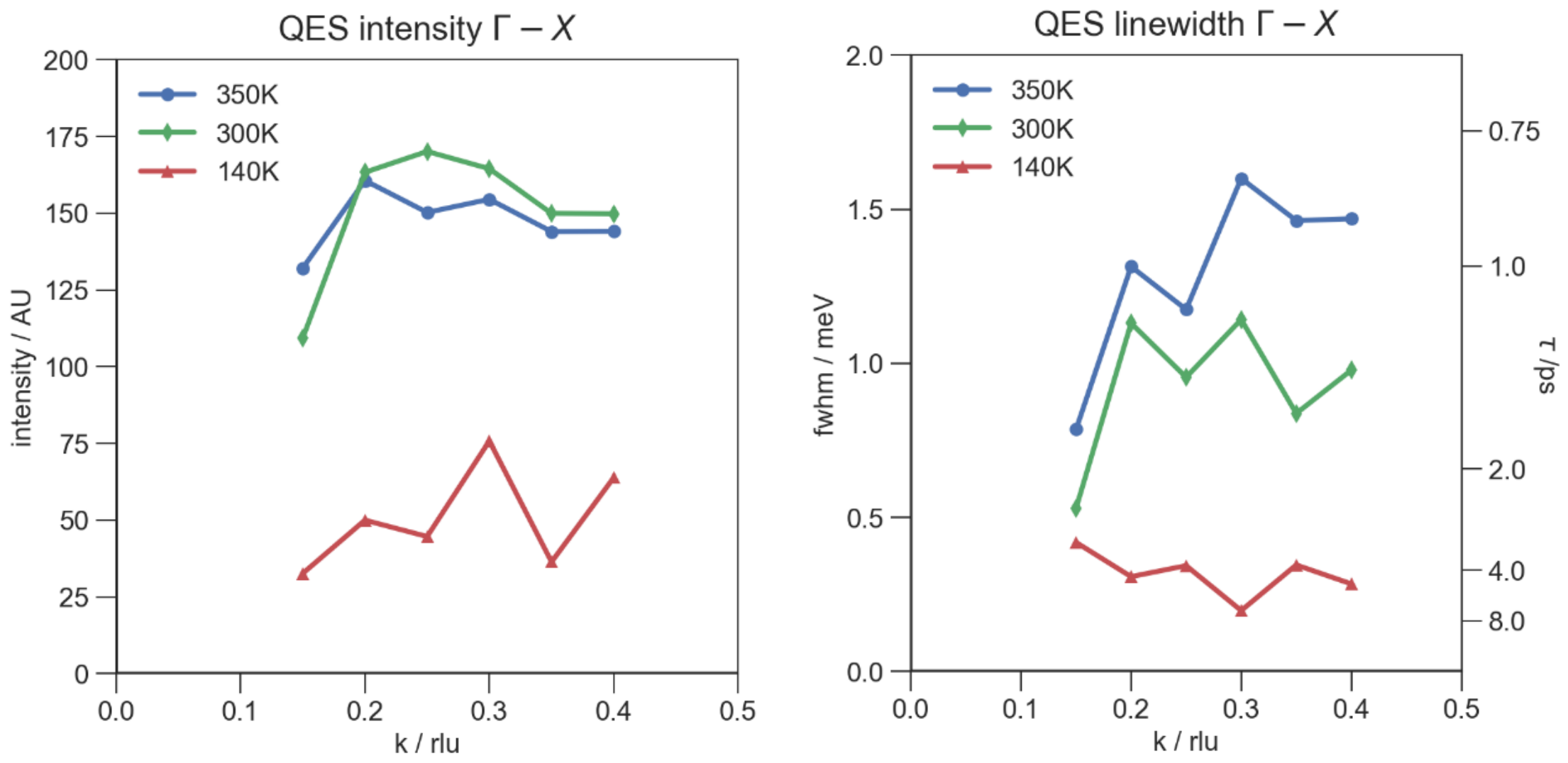

Fig. S9. QES Intensity and linewidth along $\Gamma - X$. The QES intensity increases dramatically between the orthorhombic and tetragonal phase, which we ascribe to the activation of a rotational mode in the higher-temperature phases. This rotation has a time constant of 0.8-2 ps in the high-temperature phases, which aligns with the values reported in prior studies (15, 16). Coupled with the unaltered phase transition temperatures, we take this as further indication that the lattice dynamics of deuterated MAPI are extremely similar to those in protonated MAPI.

Table S1. Instrumental energy resolution calculated for our acoustic phonon measurements as a function of reduced wavevector k (Purple lines, Fig. 4 and Fig. S5). In our fitting procedure, we fit the phonon to Voigt functions. In each constant-***Q*** scan, the gaussian width of the Voigt was fixed to the value in this table, and the Lorentzian width (the intrinsic phonon linewidth) was then determined by the fit.

| TA Γ - X | | TA Γ - M | | LA Γ - X | |
|---|---|---|---|---|---|
| k / rlu | FWHM / meV | k / rlu | FWHM / meV | k / rlu | FWHM / meV |
| 0.15 | 0.14 | 0.08 | 0.3 | 0.05 | 0.35 |
| 0.2 | 0.12 | 0.09 | 0.29 | 0.06 | 0.34 |
| 0.25 | 0.1 | 0.1 | 0.28 | 0.075 | 0.34 |
| 0.3 | 0.1 | 0.11 | 0.28 | 0.09 | 0.33 |
| 0.35 | 0.13 | 0.12 | 0.27 | 0.1 | 0.33 |
| 0.4 | 0.18 | 0.16 | 0.24 | 0.15 | 0.32 |
| | | 0.2 | 0.2 | | |
| | | 0.24 | 0.16 | | |
| | | 0.3 | 0.12 | | |
| | | 0.4 | 0.22 | | |